\documentclass[prl,
,twocolumn
,superscriptaddress,
nofootinbib,%
tightenlines ]{revtex4}
\usepackage{epsfig}
\usepackage[colorlinks=true,linkcolor=blue,urlcolor=blue,citecolor=blue]{hyperref}
\usepackage{amsmath}
\usepackage{amsfonts}
\usepackage{amssymb}

\newcommand{\ben}{\begin{displaymath}}
\newcommand{\een}{\end{displaymath}}
\newcommand{\be}{\begin{equation}}
\newcommand{\ee}{\end{equation}}
\newcommand{\bea}{\begin{eqnarray}}
\newcommand{\eea}{\end{eqnarray}}

\newcommand{\nn}{\nonumber \\ }

\begin{document}
\title{On the definition of local spatial densities in hadrons}
\author{E.~Epelbaum}
 \affiliation{Institut f\"ur Theoretische Physik II, Ruhr-Universit\"at Bochum,  D-44780 Bochum,
 Germany}
\author{J.~Gegelia}
 \affiliation{Institut f\"ur Theoretische Physik II, Ruhr-Universit\"at Bochum,  D-44780 Bochum,
 Germany}
 \affiliation{High Energy Physics Institute, Tbilisi State
University, 0186 Tbilisi, Georgia}
 \author{N.~Lange}
 \affiliation{Institut f\"ur Theoretische Physik II, Ruhr-Universit\"at Bochum,  D-44780 Bochum,
 Germany}
\author{U.-G.~Mei\ss ner}
 \affiliation{Helmholtz Institut f\"ur Strahlen- und Kernphysik and Bethe
   Center for Theoretical Physics, Universit\"at Bonn, D-53115 Bonn, Germany}
 \affiliation{Institute for Advanced Simulation, Institut f\"ur Kernphysik
   and J\"ulich Center for Hadron Physics, Forschungszentrum J\"ulich, D-52425 J\"ulich,
Germany}
\affiliation{Tbilisi State  University,  0186 Tbilisi,
 Georgia}
 \author{M.~V.~Polyakov}\thanks{Deceased.} 
  \affiliation{Institut f\"ur Theoretische Physik II, Ruhr-Universit\"at Bochum,  D-44780 Bochum,
 Germany}
\affiliation{Petersburg Nuclear Physics Institute, 
		Gatchina, 188300, St.~Petersburg, Russia}

\date{\today}
\begin{abstract}
  We show that the matrix element of a local operator between
  hadronic states gives rise to an unambiguous definition of the
  associated spatial density.
  As an explicit example,
  we consider the charge
  density of a spinless particle in the rest and moving frames and
  clarify its relationship to the electric form factor. Our results suggest
  that the interpretation of the spatial densities of local operators
  and their moments such as the mean square charge radius needs to be revised. 
\end{abstract}

\maketitle

{\it Introduction:}~It is widely accepted that the electric charge density of the nucleon
is given by the three-dimensional Fourier transform of
its
electric form factor in the Breit frame \cite{Hofstadter:1958,Ernst:1960zza,Sachs:1962zzc}.  
Similar relations have been suggested for Fourier transforms of
gravitational form factors and various local distributions in
Refs.~\cite{Polyakov:2002wz,Polyakov:2002yz,Polyakov:2018zvc}.   

The identification of spatial density distributions with the Fourier
transform of the corresponding form factors for systems whose
intrinsic size is comparable with the Compton wavelength was
criticized in
Refs.~\cite{Burkardt:2000za,Miller:2007uy,Miller:2009qu,Miller:2010nz,Jaffe:2020ebz,Miller:2018ybm,Freese:2021czn}.   
In particular, Miller pointed out that
the derivation of the conventional relationship between the charge
density and the electric form factor in the Breit frame by Sachs
\cite{Sachs:1962zzc} implicitly assumes \emph{delocalized} wave
packet states \cite{Miller:2018ybm}. This would result in moments of the charge
distribution governed by the size of the wave packet
rather than the intrinsic properties of the system encoded in the form factor. 

The definition of the charge density distribution for a spin-$0$
system was further scrutinized by Jaffe
\cite{Jaffe:2020ebz} in relationship to three characteristic
length scales: the scale $\Delta$ set by the form 
factor slope, $\Delta ^2 = 6 F' (q^2)|_{q^2 = 0}$, the characteristic size
of the wave packet $R$ and the Compton wavelength $1/m$.
Using a Gaussian wave packet and an \emph{approximate} expression for the
charge distribution, Jaffe concluded that the interpretation of the
Fourier transformed form factor as the intrinsic charge density
is not valid for light hadrons and argued that local
density distributions cannot even be defined independent of the form
of the wave packet for systems with $\Delta \sim 1/m$.   

In this paper, we revisit the definition of the charge density
for spin-$0$ systems. We closely follow the logic and conventions of
Ref.~\cite{Jaffe:2020ebz}, but we make no approximations to evaluate the
charge density in a general wave packet state. Using spherically
symmetric wave packets in the rest frame of the system, we show that
the charge density is defined unambiguously for sharply localized
packets. We then generalize the definition to moving frames and
show that  in the infinite-momentum frame (IMF), the charge density turns
into the well-known two-dimensional distribution in the transverse
plane. We also discuss the
relationship between the radial moments of the charge density and the
form factor.

{\it The charge density in the rest frame of the system:}~Following 
Ref.~\cite{Jaffe:2020ebz}, we consider, for the sake
of definiteness, a spin-$0$ system. Notice, however, that spin plays no special
role in the analysis below, which is applicable to any
localizable quantum system. We assume that the system is an eigenstate
of the charge operator $\hat Q=\int d^3{r} \, \hat \rho({\bf
  r},0)$, $\hat Q|p\rangle =Q|p\rangle$, where $\hat \rho({\bf r},0)$ is the electric charge density
operator at $t=0$ in the  Heisenberg picture, 
and we take $Q=1$ for definiteness. The momentum eigenstates
$|p\rangle$ are normalized in the usual way,
\begin{equation}
\langle p'|p\rangle = 2 E (2\pi )^3 \delta^{(3)} ({\bf p'}-{\bf p})\,,
\label{NormState}
\end{equation}
with $p=(E,{\bf p})$, $E=\sqrt{m^2+{\bf p}^2}$.
Using translati\-o\-nal invariance, the matrix elements of  $\hat \rho({\bf
  r},0)$ between momen\-tum eigenstates of a spin-$0$ system can be written as   
\begin{equation}
\langle p' | \hat\rho ({\bf r} ,0)| p \rangle  = e^{i({\bf p'}-{\bf p})\cdot {\bf r}} (E+E') F(q^2) ,
\label{eqME}
\end{equation} 
where $F(q^2)$ is the electric form factor and $q = p' - p$ denotes the
momentum transfer.

Next, we define a normalizable Heisenberg-picture state of the system
with the center-of-mass position ${\bf X}$ in terms of the wave packet
\begin{equation}
|\Phi, {\bf X}\rangle = \int \frac{d^3 {p}}{\sqrt{2 E (2\pi)^3}}  \, \phi({\bf p}) \, e^{-i {\bf p}\cdot{\bf X}} |p \rangle,  
\label{statedef}
\end{equation}
where the profile function $\phi({\bf p})$ is required to satisfy
\begin{equation}
\int d^3 {p} \,  | \phi({\bf p})|^2 =1  
\label{norm}
\end{equation}
in order to ensure the proper normalization of the wave packet. 
For later use, we define a dimensionless profile function $ \tilde
\phi$ via 
\begin{equation}
\phi({\bf p}) = R^{3/2} \, \tilde \phi(R  {\bf p})\,,
\label{packageForm}
\end{equation} 
where $R$ denotes the characteristic size of the wave packet
with $R\to 0$ corresponding to a sharp localization.
The charge density distribution in this state has the form
\begin{align}
  \rho_{\phi}({\bf X} + {\bf r}) &\equiv \langle \Phi, {\bf X} | \hat
                                   \rho ({\bf r}, 0 ) | \Phi, {\bf X}
                                   \rangle \nn
  &= \int \frac{d^3 {p} \, d^3 {p}'}{(2\pi)^3 \sqrt{4 E
    E'}}\, (E+E') \, F\left(q^2\right) \, \phi^\star({\bf p'}) \,
                                 \phi({\bf p})  \nn
  & \times e^{i {\bf q}\cdot ({\bf X} + {\bf r})},
\label{rhoint}
\end{align}
where
${\bf q}={\bf p'}-{\bf p}$ and $q^2=(E'-E)^2-{\bf q}^2$. Without loss
of generality we choose ${\bf
  X} = 0$ to place the system at the origin. Finally, introducing the
total and relative momentum variables  via
${\bf p} = {\bf P} - {\bf q}/2$ and ${\bf p}' = {\bf P} + {\bf q}/2$,
the charge density 
is written as
\begin{align}
\rho_\phi({\bf r}) &= \int \frac{d^3 {P} \, d^3 {q}}{(2\pi)^3 \sqrt{4 E E'}}\,
                (E+E') \, F\left[ (E-E')^2- {\bf q}^2\right] \nn
                &\times\phi\bigg({\bf P} -
\frac{\bf q}{2}\bigg) \, \phi^\star\bigg({\bf P} +\frac{\bf q}{2}\bigg)  \, e^{i {\bf q}\cdot {\bf  r}} ,
\label{rhoint2}
\end{align}
where the energies are $E=\sqrt{m^2+ {\bf P}^2 - {\bf P}\cdot {\bf q} +{\bf q}^2/4 } $
and $E'=\sqrt{m^2+ {\bf P}^2 + {\bf P}\cdot {\bf q} +{\bf q}^2/4 } $.

The traditional (``naive'') interpretation of the
charge density in terms of the Fourier transform of the form factor in
the Breit frame, $F(q^2) = F (-{\bf q}^2)$,
emerges by first taking the static limit (i.e., $m \to \infty$) by substituting
$E=E'=m$ in the integrand in Eq.~(\ref{rhoint2}),
\begin{align}
\rho_{\phi, \, \rm naive}({\bf r}) &= \int \frac{d^3 {P} \, d^3
                                        {q}}{(2\pi)^3}\,
                                        \phi\bigg({\bf P} - \frac{\bf
                                        q}{2}\bigg)
                                        \phi^\star\bigg({\bf P}
                                        +\frac{\bf q}{2}\bigg)  \nn
                                        &\times  F\left( -
                                        {\bf q}^2\right) \, e^{i {\bf q}\cdot {\bf r}},  
\label{rhoint2NR}
\end{align}
and subsequently localizing the wave packet by taking the limit $R \to 0$
\cite{Miller:2018ybm,Jaffe:2020ebz}. This can be done without specifying the 
functions $F\left( q^2\right)$ and $\phi ( {\bf p} )$ using the method
of dimensional counting \cite{Gegelia:1994zz} or, alternatively, the
strategy of regions \cite{Beneke:1997zp}.
For $F\left( q^2\right)$ decreasing at  large $q^2$ faster than
$1/q^2$, the only non-vanishing
contribution to $\rho_{\phi, \, \rm naive}({\bf r})$ 
in the $R\to 0$ limit is obtained by substituting ${\bf P}= \tilde {\bf
  P}/R$, expanding the integrand in Eq.~(\ref{rhoint2NR}) in $R$
around $R=0$ and keeping the zeroth order
term. The resulting naive charge density has the familiar form 
\begin{align}
\rho_{\rm naive}( r) &= \int \frac{d^3 {\tilde P} \, d^3
  {q}}{(2\pi)^3}\, F\left( - {\bf q}^2\right) \, |\tilde\phi({\tilde
  {\bf P}})|^2  \, e^{i {\bf q}\cdot {\bf  r}}\nn
&= \int \frac{d^3 {q}}{(2\pi)^3}\, F\left( - {\bf q}^2\right)  \, e^{i {\bf q}\cdot {\bf  r}},  
\label{rhoint3R}
\end{align}
where in the second equality we made use of Eq.~(\ref{norm}). Here
and in what follows, $r \equiv | {\bf r} |$.
We have dropped the subscript $\phi$ to indicate that the above
expression is independent of the
wave packet shape.

On the other hand, the method of dimensional counting
allows one to take the $R\to 0$ limit in Eq.~(\ref{rhoint2}) without
emp\-loying the static approximation. Following the same steps as
before but for
arbitrary $m$, we obtain
\begin{equation}
\rho_\phi ( {\bf r}) = \int \frac{d^3 \tilde {P} \, d^3 {q}}{(2\pi)^3}\, F\bigg[ \frac{(\tilde {\bf P}\cdot{\bf q})^2}{\tilde {\bf P}^2}- {\bf q}^2\bigg] \, |\tilde\phi(\tilde {\bf P})|^2  \, e^{i {\bf q}\cdot {\bf  r}}.  
\label{rhoint3}
\end{equation}
The resulting density depends on the
shape of the wave packet unless it is spherically symmetric. 
Since there is no preferred direction in the rest frame of the system,
we {\it define} the charge density distribution in
the rest frame by employing spherically symmetric wave
packets with $\tilde\phi(\tilde {\bf P}) =  \tilde \phi(|\tilde {\bf
  P} |)$. 
Then,  using spherical coordinates to perform the integration over
$\tilde {\bf P}$ in Eq.~(\ref{rhoint3}), we arrive at the final form
of the charge density distribution in the rest frame of a particle
\begin{equation}
\rho( r) = \int \frac{d^3 {q}}{(2\pi)^3}\, e^{i {\bf q}\cdot {\bf  r}}
\int_{-1}^{+1} d\alpha\,  \frac{1}{2}\,   F\left[ (\alpha^2-1) \,{\bf q}^2\right]\,.
\label{rhoint4}
\end{equation}

While it is argued in Ref.~\cite{Jaffe:2020ebz} that the traditional result $\rho_{\rm naive} (r)$
is valid for the hierarchy of scales $\Delta\gg 1/m$, comparing
the approximate and exact expressions in Eqs.~(\ref{rhoint3R}) and
(\ref{rhoint4}), respectively, shows that the accuracy of the static
approximation leading to $\rho_{\rm naive} (r)$ does not depend upon the particle
mass $m$. It is also clear that the validity of Eq.~(\ref{rhoint4}), which provides
an unambiguous relationship between the matrix element of the local charge
density operator $\hat \rho (0 )$ in a quantum system and the
experimentally measurable 
form factor $F (q^2)$, does not depend on the relation between the
intrinsic size of the system $\Delta$ and its Compton wavelength
$1/m$ (in contrast to what is claimed in Ref.~\cite{Jaffe:2020ebz}). 

{\it Discussion:}~A striking feature of the obtained result for 
$\rho (r)$ is its independence of the particle's mass. This implies
that the traditional expression for the charge density, $\rho_{\rm naive} (r)$, does \emph{not}
emerge from $\rho (r)$ by taking the static limit: $\rho_{\rm naive} (r) \neq \lim_{m \to
  \infty} \rho (r)$. At first glance this seems puzzling as one
expects the conventional static result
to be a better approximation for heavy systems like atoms or atomic nuclei
\cite{Miller:2018ybm,Jaffe:2020ebz}. The reason for this mismatch is
the  non-commutativity of the $R \to 0$ and $m \to \infty$ limits
of $\rho_\phi ({\bf r})$ in Eq.~(\ref{rhoint2}), as implicitly shown in 
Figs.~1-3 of Ref.~\cite{Jaffe:2020ebz}. While
the static limit and, more generally, the non-relativistic
approximation is perfectly valid when calculating the form factor in
Eq.~(\ref{eqME}) provided $-q^2 \ll m^2$, \emph{it is
violated
in certain momentum regions when performing the integration in Eq.~(\ref{rhoint2})}. 

To have a simple example demonstrating the non-com\-mu\-ta\-tivity
of the $m\to\infty$ and $R\to 0$ limits consider the wave packet in
one spatial dimension with
\begin{equation}
\phi(p) = \sqrt{\frac{2 R}{\pi}}\,\frac{1}{1+ R^2p^2} \,,
\label{defPackage}
\end{equation}
and the form factor 
\begin{equation}
F(q_0^2 - q^2) = \frac{2}{2 - \Delta^2 (q_0^2 - q^2)} \,,
\label{defFF1dim}
\end{equation}
so that $F(0)=1$ and $F' (0) = \Delta^2/2$.   
We calculate the second order moment of the charge distribution
using the version of Eq.~(\ref{rhoint2}) in one spatial dimension, 
\begin{align}
  \label{r2inOpO}
\langle x^2 \rangle_\phi &=\int_{-\infty}^{+\infty} dx \, x^2
                           \int_{-\infty}^{+\infty}
                           \frac{d P \, d  q}{ 2\pi \, \sqrt{4 E
                           E'}} (E+E') \\
                           &\times F\left[ (E-E')^2- k \, q^2 \right] 
                             \phi\bigg(P -  \frac{q}{2}\bigg)
                             \phi^\star\bigg( P +  \frac{q}{2} \bigg)   e^{i  q x } .
                             \nonumber
\end{align}
For demonstration purposes, we have introduced a control
parameter $k$  to be set to $k=1$ in the final result.
The integral in Eq.~(\ref{r2inOpO}) can be easily calculated by writing the factors of $x$
as derivatives acting on the exponential function. The resulting expression has the form
\begin{equation}
  \langle x^2 \rangle_\phi = k \Delta^2  - \frac{\Delta^2 }{(1 + m R)^2}
  + \frac{R^2}{2} - \frac{R}{4 m (1 + m R)^3}\,.
  \label{squaredradiusexact}
  \end{equation}
Taking the limit $R\to 0$ in Eq.~(\ref{squaredradiusexact}) leads to 
\begin{equation}
\langle x^2 \rangle = (k-1)\Delta^2 =0 \,,
\label{squaredradlimR} 
\end{equation}
which does not depend on the mass $m$. 
On the other hand, taking first the static limit $m\to \infty$ and
subsequently
the $R\to 0$ limit we obtain a different result
\begin{equation}
\langle x^2 \rangle_{\rm naive} =  k \Delta^2  =\Delta^2 
\,.
\label{squaredradlimRStatic} 
\end{equation}
The method of dimensional counting reproduces exactly
Eq.~(\ref{squaredradlimR}), while Eq.~(\ref{squaredradlimRStatic}) is
obtained by first taking the static limit in the integrand of
Eq.~(\ref{r2inOpO}). 

\begin{figure}[tb]
  \begin{center}
\includegraphics[width=0.42\textwidth,keepaspectratio,angle=0,clip]{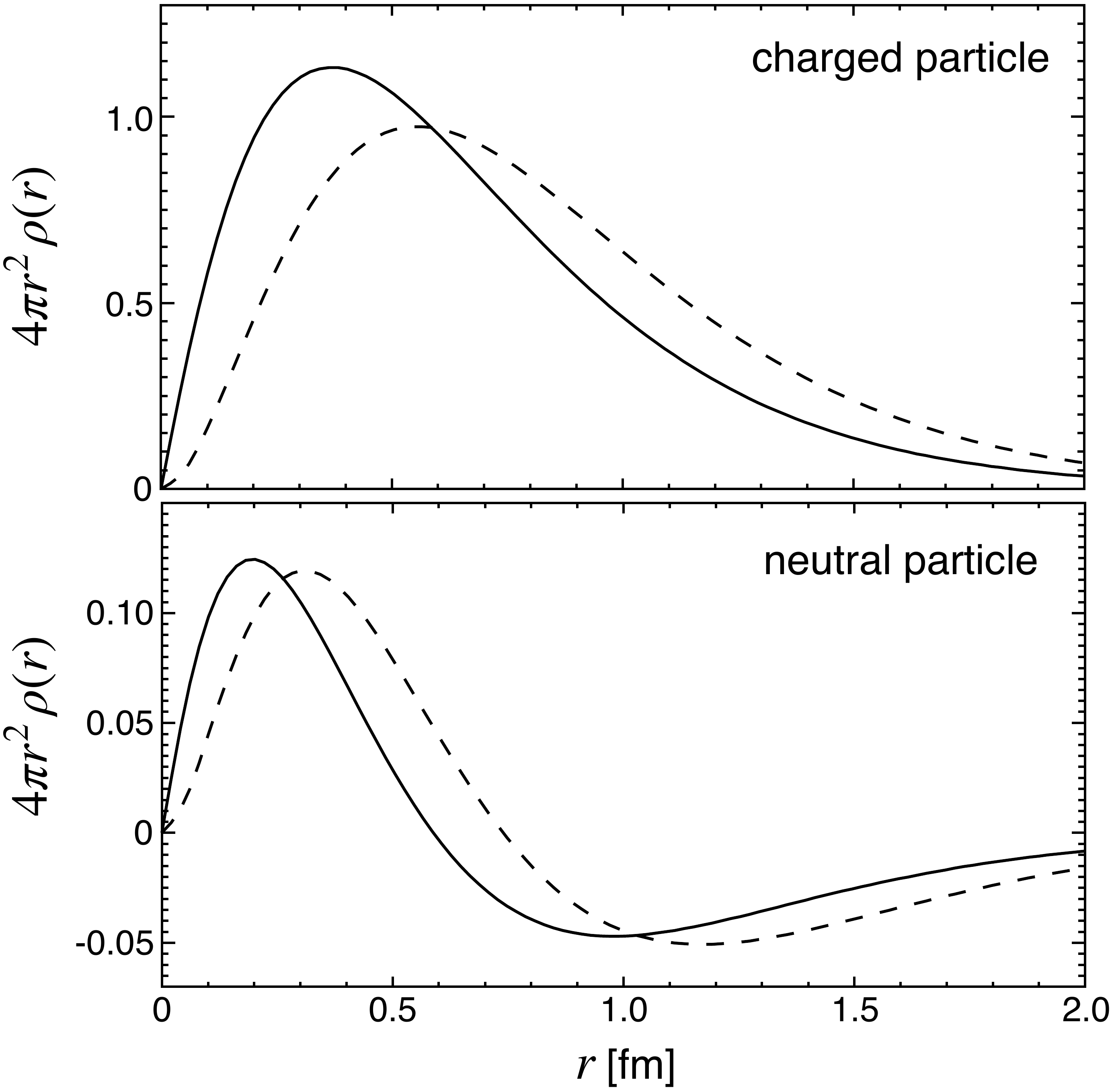}  
\end{center}
\vskip - 0.7 true cm
\caption{Radial charge density distributions $4 \pi r^2 \rho (r)$ (solid lines) and
  $4 \pi r^2 \rho_{\rm naive} (r)$ (dashed lines) for a charged and a neutral
  particle using the dipole and Galster-type parametrizations of the
  proton and neutron electric form factors, respectively. 
\label{fig1}
}
\end{figure}

We now turn to the interpretation of our result in
Eq.~(\ref{rhoint4}). Clearly, the dependence of $\rho (r)$ on the angle-averaged
form factor $\frac{1}{2} \int_{-1}^{+1} d \alpha \, F \big[
(\alpha^2 - 1) {\bf q}^2 \big]$ rather than the Breit frame
expression $F (- {\bf q}^2)$ affects the radial profile of the charge
density.
To quantify
the magnitude of this effect, we compare in Fig.~\ref{fig1} $\rho (r)$ and
$\rho_{\rm naive} (r)$ for a charged and a neutral particle. For
illustrative purposes we employ here simple parametrizations of the nucleon
form factors, namely the dipole proton form factor $F_{\rm p}
(q^2) = G_{\rm D} (q^2) = (1-q^2/\Lambda^2)^{-2}$ 
with $\Lambda^2 = 0.71$~GeV$^2$ and the Galster-type parametrization
of the neutron form factor $F_{\rm n} (q^2) = A \tau/(1 + B \tau)  \,
G_{\rm D} (q^2)$ from Ref.~\cite{Kelly:2004hm}, where $\tau = -q^2/(4 m_p^2)$, $A=1.70$ and
$B=3.30$. 

To gain further insights into
the relationship between the charge density and the form factor it is
instructive to rewrite Eq.~(\ref{rhoint4}) in coordinate independent form as 
\be
\label{RhoCoordIndep}
\rho (r) = \frac{1}{4 \pi} \int d^2 \hat n \, \rho_{\hat{\bf n}} ({\bf r})\,,
\ee
where $\hat{\bf n} \equiv {\bf n}/ | {\bf n}| $ is a unit vector and
\be
\label{RhoCoordIndep1}
\rho_{\hat{\bf n}} ( {\bf r} ) = \int \frac{d^3 q}{(2 \pi)^3} \, F (- {\bf
  q}_\perp^2) \, e^{i {\bf q} \cdot {\bf r}}
= \rho_{\hat{\bf n}} ( {r_\parallel} ) \,  \rho_{\hat{\bf n}} ( {r_\perp} )
\,.
\ee
Here, ${\bf q}_\perp = \hat{\bf n} \times ({\bf q} \times \hat{\bf n}
)$, ${\bf
  r}_\perp =\hat{\bf n} \times ({\bf r} \times \hat{\bf n}
)$, $r_\parallel = {\bf r} \cdot \hat{\bf n}$, $r_\perp \equiv |{\bf  r}_\perp |$,
and the
one- and two-dimensional densities in the $\hat{\bf
  n}$ and ${\bf
  r}_\perp $ directions are given by 
\begin{align}
 \label{RhoCoordIndep2} 
  \rho_{\hat{\bf n}} ( r_\parallel ) &=\int
                                       \frac{d q_\parallel}{2 \pi }
                                       \, e^{i {q}_\parallel  
                                       {r}_\parallel} = \delta (
                                     r_\parallel )\,, \\
\rho_{\hat{\bf n}} ( {r}_\perp ) &=         \int
                                       \frac{d^2 q_\perp}{(2 \pi )^2}
                                       \, F  (- {\bf
                                       q}_\perp^2) \, e^{i {\bf q}_\perp \cdot {\bf r}_\perp}\,,
                                       \nonumber
\end{align}
with $q_\parallel = {\bf q} \cdot \hat{\bf n}$.
These expressions
make it clear that $F (q^2)$ effectively measures the deviation from the
point-like limit
in \emph{two} rather than three spatial
dimensions. This in turn explains the squeezing of the radial charge
density relative to the naive result as shown in Fig.~\ref{fig1}.  

To further elaborate on this point we compute
radial moments of the charge distributions in $d \in\mathbb{N}$
spatial dimensions.
We start with the inverse Fourier transform of Eq.~(\ref{rhoint3R})
\be
F (- {\bf q}^2) = \int d^d r \, \rho_{\rm
  naive} (r) \, e^{- i {\bf q} \cdot {\bf r}}\,.
\ee
Taking the $k^{\rm th}$ derivative of this expression at $- {\bf q}^2=0$,
$F^{(k)} (0)$, we find
for $\langle r^{2k} \rangle_{\rm naive}^{(d)} \equiv \int d^d{r} \, \rho_ {\rm
  naive} ({r}) \,  r^{2k}$: 
\be
\label{MomNaiveD}
\langle r^{2k} \rangle_{\rm naive}^{(d)} = \frac{2^{2k} \, \Gamma
  (d/2+k)}{\Gamma (d/2)} \, F^{(k)} (0)\,.
\ee
For $d=3$, this reduces to the well-known expression
\be
\label{MomNaive3}
\langle r^{2k} \rangle_{\rm naive} = \frac{(2k+1)!}{k!} F^{(k)} (0)\,.
\ee

On the other hand, using
Eqs.~(\ref{RhoCoordIndep})-(\ref{RhoCoordIndep2}) generalized to $d$
dimensions and noting that
$\langle r^{2k} \rangle_{\hat{\bf n}}^{(d)} \equiv \int d^dr \, \rho_{\hat{\bf
    n}} ({\bf r})\, r^{2k}$ does not depend of $\hat{\bf n}$, we obtain
for $d \ge 2$
\be
\label{MomD}
\langle r^{2k} \rangle^{(d)} = \langle r^{2k} \rangle_{\rm naive}^{(d-1)} \,,
\ee
so that in $d=3$ spatial dimensions,
\be
\label{Mom3}
\langle r^{2k} \rangle = 2^{2k} \, k! \, F^{(k)} (0)\,.
\ee
Notice further that in one spatial dimension, $\rho (r) = \delta (r)$
independently of the form factor. This explains the vanishing result
for the second moment in the considered one-dimensional example, see
Eq.~(\ref{squaredradlimR}).

{\it The charge density in moving frames:}~While the sta\-tic
approximation
$\rho_{\rm naive} (r)$ does not depend on the frame, the expressions for $\rho (r)$ in Eqs.~(\ref{rhoint4})
and (\ref{RhoCoordIndep}) are valid in the rest frame of the system.
It is straightforward to generalize these results to a boosted frame.

We start with the general expression for $\rho_\phi ({\bf r})$ in
Eq.~(\ref{rhoint2}) and replace $\phi ({\bf p} )$ with $\phi_{\bf v}
({\bf p} )$, where ${\bf v}$ denotes the boost velocity. Differently to
$\phi ({\bf p} )$, we cannot regard the function  $\phi_{\bf v}
({\bf p} )$ to be spherically symmetric. Thus, we need to express 
$\phi_{\bf v} ({\bf p} )$ in terms of the rest frame quantity $\phi
({\bf p} )$
in order to obtain a wave packet independent definition for the charge
density in the $R\to 0$ limit.  Using Eq.~(\ref{statedef}) and the Lorentz transformation
properties of the momentum eigenstates $| p \rangle$, one finds
\cite{Hoffmann:2018edo}
\be
\label{temp01}
\phi_{\bf v} ({\bf p})= \sqrt{\gamma \Big( 1-\frac{{\bf v} \cdot {\bf
      p}}{E} \Big)} \, \phi \big[ {\bf p}_\perp + \gamma ( {\bf
  p}_\parallel - {\bf v} E )\big]\,,
\ee
where $\gamma = (1 - v^2)^{-1/2}$, ${\bf p}_\parallel = ({\bf p} \cdot
\hat {\bf v}) \hat {\bf v}$, ${\bf p}_\perp = {\bf p} - {\bf p}_\parallel$ and
$E = \sqrt{m^2 + {\bf p}^2}$. We note in passing that
Eq.~(\ref{temp01})
ensures the invariance of the normalization of
the wave packet \cite{Hoffmann:2018edo}.
Then, following the same steps as 
in the case of the rest frame and using the method of dimensional counting to
evaluate the $R \to 0$ limit we arrive at 
\begin{align}
\rho_{\phi,  {\bf v}} ( {\bf r}) &= \int \frac{d^3 \tilde {P} \, d^3
  {q}}{(2\pi)^3}\,
\frac{\gamma (\tilde {P}  - {\bf v} \cdot \tilde {\bf P})}{\tilde {P}}
\, 
F\bigg[ \frac{(\tilde {\bf P}\cdot{\bf q})^2}{\tilde {\bf P}^2}- {\bf
                                   q}^2\bigg] \nn
    &\times                                
      \, \Big|\tilde\phi \big[
  \tilde {\bf P}_\perp + \gamma (  \tilde {\bf P}_\parallel - {\bf v}
      \tilde {P} ) \big] \Big|^2  \, e^{i {\bf q}\cdot {\bf  r}}\,,
\label{rhoint3boosted}
\end{align}
where $\tilde {P} \equiv | \tilde {\bf P}|$. We now change the
integration variable $ \tilde {\bf P} \to \tilde {\bf P}' =  \tilde
{\bf P}_\perp + \gamma (  \tilde {\bf P}_\parallel - {\bf v}
      \tilde {P} )$. Using the relations $\tilde {\bf P}_\parallel   = \gamma ( \tilde {\bf P}_\parallel ' + {\bf
                             v}  \tilde {P}' )$ and $\tilde { P}       = \gamma ( \tilde {P}' + {
                             v}  \tilde {P}_\parallel ')$, 
it is easy to verify
that the Jacobian of the change of variables $\tilde {\bf
  P} \to \tilde {\bf P}'$ cancels the first factor in the
integrand in Eq.~(\ref{rhoint3boosted}), yielding
\begin{align}
  \label{rhoint3boosted2}
\rho_{\phi,  {\bf v}} ( {\bf r}) &= \int \frac{d^3 \tilde {P}' \, d^3
  {q}}{(2\pi)^3}
                                   \; \big|\tilde\phi \big(
\tilde {\bf P} ' \big) \big|^2  \, e^{i {\bf q}\cdot {\bf  r}} \\
              &\times                     F\left\{ \frac{\big[\tilde {\bf P}_\perp
                                   ' \cdot {\bf q}_\perp + \gamma ( \tilde {\bf P}_\parallel ' + {\bf
                             v}  \tilde {P}' ) \cdot {\bf q}_\parallel
                                   ]^2}{\gamma^2 ( \tilde {P}' + {
                             v}  \tilde {P}_\parallel ')^2} - {\bf
                q}^2 \right\} .
                \nonumber
\end{align}
Using Eq.~(\ref{norm}) and the spherical symmetry of $\tilde\phi \big(
\tilde {\bf P} ' \big)$, the integration over $\tilde P'$ becomes trivial.
The remaining angular integration over $\hat {\tilde {\bf P}}'$ can be done in
spherical coordinates. We align the $z$- and $x$-axes along the ${\bf v}$
and ${\bf q}_\perp$ directions, respectively, and denote $\eta =
\cos \theta $. Our final result then reads:
\be
\label{RhoBoostedCoord1} 
\rho_{\bf v} ({\bf r})= \int \frac{d^3 {q}}{(2\pi)^3}\, \bar F\left(
  q_\parallel,   {q}_\perp \right)  \, e^{i {\bf q}\cdot {\bf  r}}\,,
\ee
with $q_\parallel \equiv \hat{\bf v} \cdot {\bf q}$, $q_\perp \equiv | {\bf q}_\perp |$ and 
\begin{align}
 \label{RhoBoostedCoord2} 
\bar F (
  q_\parallel,   {q}_\perp ) &=\frac{1}{4 \pi} \int_{-1}^{+1}
d\eta \int_0^{2 \pi} d\phi \\
& \times 
F\left\{ \frac{\big[\sqrt{1-\eta^2} \cos \phi \, q_\perp + \gamma (\eta + v)
                                 q_\parallel \big]^2}{\gamma^2 (1 + v
                                 \eta)^2} - {\bf q}^2 
                                 \right\}.
\nonumber
\end{align}
In the IMF with $v \to 1$ and
$\gamma \to \infty$, the charge density turns into the usual
two-dimensional distribution in the transverse plane, $\rho_{\rm IMF}
({\bf r}) = \delta (r_\parallel ) \, \rho_{\rm IMF} (r_\perp )$ with
\be
\label{IMF}
\rho_{\rm IMF} (r_\perp )= \int \frac{d^2 {q}_\perp}{(2\pi)^2}\, F\left(
  -{\bf q}_\perp^2 \right)  \, e^{i {\bf q}_\perp \cdot {\bf  r}_\perp}\,.
\ee
One can also verify that Eq.~(\ref{RhoBoostedCoord1}) reduces to the
rest frame expression in Eq.~(\ref{rhoint4}) in the limit $v \to 0$,
albeit this relationship appears somewhat obscured.

Again, it is instructive to rewrite Eqs.~(\ref{RhoBoostedCoord1},\ref{RhoBoostedCoord2})
in a coordinate independent form similar to the rest frame expressions in
Eqs.~(\ref{RhoCoordIndep})-(\ref{RhoCoordIndep2}). We introduce a unit
vector $\hat {\bf m} \equiv \hat{\tilde {\bf P}}'$ and define a vector
valued function
\be
{\bf n} ({\bf v}, \hat {\bf m}) = \hat {\bf v} \times (\hat {\bf m}
\times \hat {\bf v} ) + \gamma (\hat {\bf m} \cdot \hat {\bf v} + v ) \hat {\bf v}\,.
\ee
Then, the charge density $\rho_{\bf v} ({\bf r})$ can be written as
\be
\label{RhoBoostedCoordIndep}
\rho_{\bf v}  ({\bf r}) = \frac{1}{4 \pi} \int d^2 \hat m \, \rho_{\hat{\bf n}({\bf v}, \hat {\bf m})} ({\bf r})\,,
\ee
where $\rho_{\hat{\bf n}({\bf v},
  \hat {\bf m})} ({\bf r}) \equiv \rho_{\hat{\bf n}} ({\bf r})$ is defined in Eqs.~(\ref{RhoCoordIndep1}), (\ref{RhoCoordIndep2}).
In this form, both extreme limits for the boosting velocity become
particularly transparent by using the relations $\hat {\bf n} ({\bf v},
\hat {\bf m}) \stackrel{v \to 0}{\longrightarrow} \hat {\bf m}$ and 
$\hat {\bf n} ({\bf v},
\hat {\bf m}) \stackrel{v \to 1}{\longrightarrow} \hat {\bf v}$,
leading evidently to Eqs.~(\ref{RhoCoordIndep}) and (\ref{IMF}),
respectively.

Last but not least, we emphasize that radial moments of the charge
distribution are, in fact, frame independent, i.e.~$\langle r^{2k}
\rangle_{\bf v}=\langle r^{2k}
\rangle$, in spite of $\rho_{\bf v}
({\bf r} )$ being not spherically symmetric for $v \neq 0$. This
remarkable feature follows from Eq.~(\ref{RhoBoostedCoordIndep}) by noting that $\int
d^3r \, \rho_{\hat{\bf n}({\bf v}, \hat {\bf m})} ({\bf r})\, r^{2k}$
does not depend on ${\bf v}$ and $\hat {\bf m}$. It can also be
verified by showing that radial moments of $\rho_\phi ({\bf r})$ in
Eq.~(\ref{rhoint3}) do not depend on $\tilde \phi (\tilde {\bf P} )$ even if
this function is not spherically symmetric. 

{\it Summary and conclusions:}~In summary, we introduced an unambiguous definition of a spatial
distribution of the expectation values of local operators in spin-$0$ systems independent of the
specific form of the wave packet in which the state was prepared.
Our definition also applies to systems whose intrinsic size is
comparable or even smaller than the Compton wavelength. 
We found remarkably simple relationships between
the electric form factor and the charge density in the rest and moving
frames, thereby reproducing the well-known result in the infinite
momentum frame. We have also demonstrated that radial moments of the charge
distribution are frame independent.

Our results suggest that form factors effectively measure the 
deviation from the point-like limit in two rather than three spatial
dimensions. This implies, in particular, that the second moment of the
charge distribution, the quantity that should be interpreted as the
mean square charge radius of the system, is related to the form factor
slope via $\langle r^2 \rangle = 4 F' (0)$ in contrast to the usual
relationship $\langle r^2 \rangle_{\rm naive} = 6 F' (0)$ motivated by
the Breit frame distribution $\rho_{\rm naive} (r)$. Thus, the actual
size of e.g.~the proton measured by the charge distribution
is $\sqrt{\langle r_{\rm p}^2 \rangle} =
0.6866(3)$~fm rather than  $\sqrt{\langle r_{\rm p}^2 \rangle_{\rm naive}} =
0.8409(4)$~fm \cite{ParticleDataGroup:2020ssz} as commonly accepted. Differently to
Refs.~\cite{Burkardt:2000za,Miller:2007uy,Miller:2009qu,Miller:2010nz,Jaffe:2020ebz,Miller:2018ybm,Freese:2021czn},
our results show that the approximation $\rho_{\rm naive} (r)$ does
not emerge in the static limit of the exact expression for $\rho (r)$, and its
accuracy is independent of the particle's mass.

We note that for heavy systems
with $(m \Delta )^{-1} \equiv \epsilon  \ll 1$, one may alternatively attempt to
define $\rho ({\bf r})$ using wave packets with $\epsilon \ll  (m R)^{-1} \ll 1$ as
suggested in \cite{Jaffe:2020ebz}, by choosing e.g.~$(m R)^{-1} \sim
\mathcal{O} (\epsilon^{1/2})$. While this leads to an unambiguous definition
of the charge density in the static limit $\epsilon \to 0$ with
$\rho ({\bf r}) \to \rho_{\rm naive} (r)$, corrections beyond this limit
are wave packet dependent. 

Our analysis can be straightforwardly generalized to
systems with non-vanishing spin and to spatial distributions introduced in
Refs.~\cite{Hofstadter:1958,Ernst:1960zza,Sachs:1962zzc,Polyakov:2002wz,Polyakov:2002yz,Polyakov:2018zvc}.  

\bigskip
{\it Acknowledgements:}~We thank Bob Jaffe for useful comments on the
manuscript.
This work was supported in part by BMBF (Grant No. 05P18PCFP1), by
DFG and NSFC through funds provided to the Sino-German CRC 110
“Symmetries and the Emergence of Structure in QCD” (NSFC Grant
No. 11621131001, DFG Project-ID 196253076 - TRR 110),
by ERC  NuclearTheory (grant No. 885150) and ERC EXOTIC (grant No. 101018170),
by CAS through a President’s International Fellowship Initiative (PIFI)
(Grant No. 2018DM0034), by the VolkswagenStiftung
(Grant No. 93562), by the EU Horizon 2020 research and
innovation programme (STRONG-2020, grant agreement No. 824093),
and by the Heisenberg-Landau Program 2021.

\end{document}